\documentclass[aps,pra,twocolumn,showpacs,amsmath,amssymb]{revtex4}

\usepackage{graphicx} 
\usepackage{dcolumn} 
\usepackage{bm} 
\usepackage{epsfig}
\usepackage{amsmath}
\usepackage{amssymb}
\usepackage{hyperref}
\begin{document}
\newcommand{\bsy}[1]{\mbox{${\boldsymbol #1}$}} 
\newcommand{\bvecsy}[1]{\mbox{$\vec{\boldsymbol #1}$}} 
\newcommand{\bvec}[1]{\mbox{$\vec{\mathbf #1}$}} 
\newcommand{\btensorsy}[1]{\mbox{$\tensor{\boldsymbol #1}$}} 
\newcommand{\btensor}[1]{\mbox{$\tensor{\mathbf #1}$}} 
\newcommand{\tensorId}{\mbox{$\tensor{\mathbb{\mathbf I}}$}} 
\newcommand{\be}{\begin{equation}}
\newcommand{\ee}{\end{equation}}
\newcommand{\bea}{\begin{eqnarray}}
\newcommand{\eea}{\end{eqnarray}}
\newcommand{\e}{\mathrm{e}}
\newcommand{\arccot}{\mathrm{arccot}}
\newcommand{\arctanh}{\mathrm{arctanh}}

\title{Transmission of electromagnetic waves through nonlinear optical layers}

\author{L. M. Hincapie-Zuluaga, J. D. Mazo-V\'asquez, C. A. Betancur-Silvera, and E. Reyes-G\'omez}
\affiliation{Instituto de F\'{\i}sica, Universidad de Antioquia UdeA, Calle 70 No. 52-21, Medell\'{\i}n, Colombia}

\begin{abstract}
The excitation of soliton states in optical layers exhibiting Kerr nonlinearities is theoretically investigated. The optical transmission coefficient is obtained as a function of the intensity of an incident light beam. The incident electromagnetic wave is considered to be monochromatic and linearly polarized in the transverse electric configuration. In materials exhibiting self-defocusing nonlinearity, soliton excitations are not observed if the product of the nonlinear dielectric coefficient of the slab and the absolute square of the incident electric-field amplitude is below a well-defined value. At such a limit, a soliton excitation with a position-independent electric field amplitude is observed within the nonlinear layer, regardless of the frequency value of the incident wave.
\end{abstract}


\date{\today}

\maketitle

The propagation of electromagnetic waves through nonlinear materials has been extensively investigated over the last decades. The nonlinear response of a medium, i.e., the dependence of the system's optical properties on the intensity of the incident electromagnetic field, leads to the existence of physical phenomena that are not observed in conventional linear materials. Different mechanisms could give rise to nonlinear optical properties in various materials. For example, it has been shown that thermal effects and induced polarization could be responsible for third-order optical nonlinearity in high-purity silica samples containing Cu nanoparticles \cite{TorresJAP2008}. Changes in molecular orientation in organic solvents and solutions could lead to the nonlinear behavior of the refractive index \cite{BundulisJOSAB2020}, etc. Of particular interest turns out to be the excitation of transparency states in low-absorptive nonlinear materials when the intensity of the incident electromagnetic radiation is appropriately tuned. In such cases, the system is observed to switch from states of low transparency to states of maximum transmission, a situation that can be understood from the concept of soliton \cite{Lomdahl1984}. A change in the incident electromagnetic-field intensity can cause soliton excitations, propagating through the nonlinear medium with a maximum transmission. This phenomenon, known as transmission-switching phenomenon, has been extensively studied in both nonlinear thin films \cite{ChenPRB1987-1,PeschelJOSAB1988, LeungPRB1989} and multilayer structures built by alternating linear and nonlinear materials along a predefined growth axis \cite{TrutschelJOSAB1988,ChenPRB1987-2}. In this particular case, soliton excitation was reported for frequency values in the vicinity of the Bragg gaps \cite{ChenPRL1987}. Moreover, when the linear material composing the periodic structure exhibits left-handed material properties, soliton excitations associated with both the zero-n gap \cite{HegdeOL2005} and the longitudinal plasmon-polariton gap \cite{CavalcantiOL2014} can also be observed.

Of great relevance for the present study are the works by Chen and Mills \cite{ChenPRB1987-1} and Leung \cite{LeungPRB1989}, who were able to develop exact analytical approaches to theoretically investigate the tunneling effect of electromagnetic waves through non-absorptive Kerr-like nonlinear layers. They also investigated the transmission-switching phenomenon in both self-focusing and self-defocusing materials. Despite the considerable amount of literature on the formation of solitons in nonlinear layers, a detailed review of the works mentioned above has led us to understand some transmission properties of nonlinear systems that, at the time, were not clearly explained. The present work aims to investigate such properties. In particular, we will show that, for soliton excitations to occur in layers with a self-defocusing Kerr nonlinearity, the product of the nonlinear dielectric coefficient of the slab and the absolute square of the incident electric-field amplitude must exceed a specific minimum value. This finding opens up the possibility of predicting intensity regions where the formation of solitons would take place and other ones where such effects would not be observed at all, and could be of relevance in optical-computing investigations \cite{CorbelliJIS2013,SilvaNJP2021}.

\begin{figure}
\centering
\epsfig{file=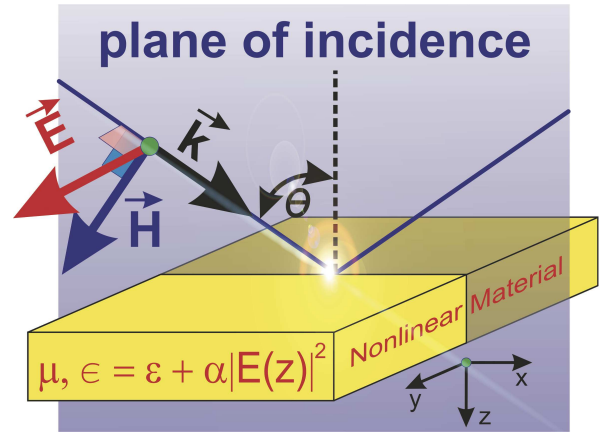,width=0.9\columnwidth}
\caption{(Color online) Pictorial view of an electromagnetic wave incident on a nonlinear layer, with the angle of incidence $\theta$ and electric-field intensity $\bvec{E}$. The electromagnetic response of the nonlinear layer is characterized by the magnetic permeability $\mu$ and the electric permittivity $\epsilon$. The corresponding parameters in the surrounding linear medium are $\mu_0$ and $\varepsilon_0$, respectively. The incident electromagnetic wave is assumed to be polarized in the transverse electric configuration. The plane of incidence is the $xz$ plane containing the wave vector $\bvec{k}$, and the electric field is oriented along the $y$ direction.}
\label{fig1}
\end{figure}

We consider an electromagnetic wave of frequency $\omega=2 \pi \nu$ incident on a slab of thickness $d$, electric permittivity $\epsilon$ and magnetic permeability $\mu$, grown along the $z$ direction as shown in Fig. \ref{fig1}. We suppose that the slab is composed of a non-absorptive material, that is, both $\epsilon$ and $\mu$ are real parameters. The incidence angle, denoted as $\theta$, is measured in the surrounding medium where the electric permittivity and magnetic permeability are given by $\varepsilon_0$ and $\mu_0 $, respectively. The wave is assumed to be linearly polarized according to the transverse electric (TE) configuration, i.e., the electric-field intensity of the incident wave is perpendicular to the plane of incidence. The electric-field component of the electromagnetic wave is then given by $\bvec{E} (\bvec {r}, t) = {\cal E} (z) e^{i (qx- \omega t)} \bvec {e}_y $, where $q = \frac {\omega} {c} \sqrt {\varepsilon_0 \mu_0} \sin (\theta)$ is the projection of the wave vector $\bvec{k}$ along the $x$ direction (cf. Fig. \ref{fig1}) and $\bvec {e}_y$ is the unit vector along the $y$ direction. The electric-field amplitude of the wave satisfies the equation \cite{CavalcantiOL2014}
\be
\label{eq1}
-\frac{d}{dz}\frac{1}{\Bar{\mu}(z)}\frac{d}{dz} {\cal E}(z) = \left[ \frac{\omega^2}{c^2} \Bar{\varepsilon}(z) - \frac{\omega^2}{c^2 \Bar{\mu}(z)}\mathrm{sin^2} (\theta) \right]{\cal E}(z),
\ee
where $\Bar{\varepsilon}(z)$ and $\Bar{\mu}(z)$ are the position-dependent electric permittivity and magnetic permeability, respectively, along the $z$ growth axis. One may notice that $\Bar{\varepsilon}(z) = \varepsilon_0$ and $\Bar{\mu}(z) = \mu_0$ in the surrounding medium, whereas $\Bar{\varepsilon}(z) = \epsilon$ and $\Bar{\mu}(z) = \mu$ within the slab.

We assume that the surrounding medium is linear and non-dispersive, whereas the optical material composing the slab exhibits a Kerr-type nonlinear behavior. The nonlinear electric permittivity is then related to the third-order susceptibility tensor, resulting from the third-order expansion of the nonlinear polarization in series of powers of the electric field. Here we suppose that the optical material is isotropic. Therefore, as the incident electromagnetic wave is considered to be linearly polarized, the electric permittivity of the slab may be given by \cite{Sutherland2003} $\epsilon = \varepsilon + \alpha \vert {\cal E} \vert^2$, where $\varepsilon$ is the dielectric background constant of the slab. The parameter  $\alpha$ is the so-called nonlinear dielectric coefficient which is a real-scalar number proportional to the real part of the $\chi^{(3)}_{yyyy}$ component of the third-order susceptibility tensor \cite{Sutherland2003,Kivshar2003} of the nonlinear layer. The term $\alpha \vert {\cal E} \vert^2$ is the nonlinear Kerr contribution to the electric permittivity. It is important to note that the sign of $\alpha$ is determined by the sign of $\chi^{(3)}_{yyyy}$ \cite{Kivshar2003}. The case $\alpha>0$ corresponds to self-focusing materials, whereas $\alpha<0$ corresponds to self-defocusing nonlinearities. We want to emphasize that our mathematical derivation below is consistent with any value of the sign of $\alpha$.

The optical parameters that describe the system depend, in general, on the wave frequency. For instance, it has been shown in recent experimental work that simultaneous changes in wave frequency and slab thickness lead to switching between self-focusing and self-defocusing regimes in certain nonlinear materials under femtosecond and picosecond laser pulses \cite{TorresOE2013}. However, in other experimental situations, considering the optical parameters independent of the wave frequency can be a good approximation \cite{PeschelJOSAB1988}. For the sake of simplicity, we have taken $\varepsilon_0$, $\mu_0$, $\varepsilon$, $\mu$, and $\alpha$ as frequency-independent parameters, an approximation that might be valid in specific experimental situations.

One may propose
\be
\label{eq2}
{\cal E}(z) = E_i
\begin{cases}
e^{i Q_0 z} + r e^{-i Q_0 z} & $if $ z<0 \cr
\phi(z) & $if $ 0<z<d \cr
t e^{i Q_0 (z-d)} & $if $ z>d
\end{cases}
\ee
as the solution of Eq. \eqref{eq1}, where $E_i$ is the amplitude of the incident electric field, $Q_0 = \frac{\omega}{c} p_0$, and $p_0 = \sqrt{\varepsilon_0 \mu_0} \cos {(\theta)}$. By taking into account the continuity of ${\cal E}$ and $\frac{1} {\Bar{\mu}} \frac{d} {dz} {\cal E}$ at the interfaces, one may find the reflection ($R = rr^*$) and transmission ($T = tt^*$) coefficients corresponding to the slab. By defining $\zeta = \frac{\omega}{c} z$, Eq. \eqref{eq1} becomes
\be
 \label{eq3}
\frac{d^2}{d\zeta^2} \phi(\zeta) + \left( \varepsilon \mu - \varepsilon_0 \mu_0 \mathrm{sin^2}\theta \right)\phi(\zeta) + a \mu \vert \phi(\zeta) \vert^2 \phi(\zeta) = 0
\ee
within the region $0 < \zeta < \zeta_d$, where $\zeta_d = \frac{\omega}{c}d$. We refer to $a = \alpha \vert E_i\vert^2$ as the normalized input intensity  \cite{PeschelJOSAB1988} since such parameter is proportional to the intensity of the incident wave but could be either positive or negative, depending on the sign of the nonlinear dielectric coefficient corresponding to the material layer. We propose $\phi = f \mathrm{e}^{i\varphi}$, where $f$ and $\varphi$ are real functions of $\zeta$. Accordingly, Eq. \eqref{eq3} leads to
\be
\label{eq4}
\frac{d}{d\zeta} \left( f^2 \frac{d\varphi}{d\zeta}\right) = 0,
\ee
and 
\be
\label{eq5}
\frac{d^2f}{d\zeta^2}- f\left(\frac{d\varphi}{d\zeta}\right)^2 + p^2f + a \mu f^3 = 0,
\ee
where $p^2 = \varepsilon \mu - \varepsilon_0 \mu_0 \sin^2 (\theta)$. The first integrals of Eqs. \eqref{eq4} and \eqref{eq5} are
\be 
\label{eq6}
\frac{d\varphi}{d\zeta} = \frac{A}{f^2}
\ee
and
\be 
\label{eq7}
\left( \frac{1}{2} \frac{dF}{d\zeta}\right)^2 + A^2 + p^2 F^2 + \frac{a\mu}{2}F^3 - BF = 0,
\ee
respectively, where $A$ and $B$ are integration constants and $F = f^2$. We calculate the integration constants $A$ and $B$ by taking into account the boundary conditions that the electric-field amplitude fulfills at the slab interfaces. From the continuity of $\phi$ and $\frac{1} {\mu} \frac{d\phi} {d \zeta}$ at the interface $\zeta=0$ one finds
\be
\label{eq8}
\begin{cases}
1 - r = \frac{1}{i\eta}\left[\dot{f}_0 \mathrm{e}^{i\varphi_0} + i\dot{\varphi}_0 f_0 \mathrm{e}^{i\varphi_0} \right] \cr
1 + r = f_0 \mathrm{e}^{i\varphi_0}, 
\end{cases}
\ee
where $\eta = p_0 \frac{\mu}{\mu_0}$, $f_0 = f(0), \varphi_0 = \varphi(0)$, $\dot{f}_0 = \frac{df}{d\zeta}\vert_{\zeta=0}$, and $\dot{\varphi}_0 = \frac{d\varphi}{d\zeta}\vert_{\zeta=0}$. One may note, from the above equations, that
\be
\label{eq9}
4\eta^2 = \dot{f}_0 ^2 + f_0^2(\eta + \dot{\varphi}_0)^2.
\ee
Similarly, at the interface $\zeta = \zeta_d$ one obtains
\be
\label{eq10}
\begin{cases}
t = f_d \mathrm{e}^{i\varphi_d} \cr
t = \frac{1}{i\eta}\left[\dot{f}_d \mathrm{e}^{i\varphi_d} + if_d \dot{\varphi}_d \mathrm{e}^{i\varphi_d}\right],
\end{cases}
\ee
where $f_d = f(\zeta_d), \varphi_d = \varphi(\zeta_d)$, $\dot{f}_d = \frac{df}{d\zeta}\vert_{\zeta=\zeta_d}$, and $\dot{\varphi}_d = \frac{d\varphi}{d\zeta}\vert_{\zeta=\zeta_d}$. Notice that the transmission coefficient is given by $T= tt^* = f_d^2 = F_d$. From the set of Eqs. \eqref{eq10} one may see that $\dot{\varphi}_d = \eta$ and $\dot{f_d} = 0$. By evaluating Eq. \eqref{eq6} at the interface $\zeta = \zeta_d$ one obtains $A = \eta \,F_d$. Moreover, from Eq. \eqref{eq7} evaluated at $\zeta = \zeta_d$ one has the constant $B = F_d \left [ \eta^2 + p^2 + \frac{a \mu}{2} F_d \right]$ as a function of both $F_d$ and $a$. Finally, we use the above results to expand Eq. \eqref{eq9}. In the process we have taken into account that $\dot{\varphi}_0 = \frac{A}{F_0}$ [cf. Eq. \eqref{eq6}], where $F_0 = f_0^2$. The result is
\be
\label{eq11}
4 \eta^2 - F_d \left [ 3 \eta^2 + p^2 + \frac{a \mu}{2} F_d \right ] + F_0 \left [ p^2 - \eta^2 + \frac{a \mu}{2} F_0 \right ] = 0.
\ee

\begin{figure}
\centering
\epsfig{file=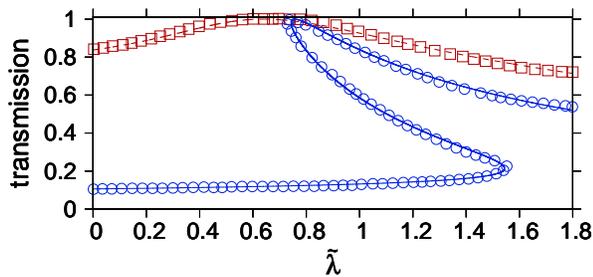, width=0.9\columnwidth}
\caption{(Color online) Transmission coefficient as a function of $\tilde{\lambda} = \frac{\alpha} {\varepsilon} \vert E_i \vert^2$. Lines correspond to the results obtained from Eq. \eqref{eq11}, whereas symbols correspond to theoretical results reported in Ref.  \cite{ChenPRB1987-1}. Calculations were performed for $\varepsilon_0 = \mu_0 = 1$, $\mu = 1$, and $\alpha>0$. The dashed line and the squares correspond to $\varepsilon = 4$, whereas the circles and the solid line correspond to $\varepsilon = 100 $. In each case, the thickness of the slab was taken as $d = 0.4 \lambda_f$, where $\lambda_f = \frac{2 \pi}{\omega} \frac{c}{\sqrt{\varepsilon \mu}}$.}
\label{fig2}
\end{figure}

The above equation may be used to obtain the transmission coefficient as a function of $a = \alpha \vert E_i \vert^2$. The procedure consists of providing a value of $F_d$ such that $0 < F_d \leq 1$, that is, the transmission coefficient $T$. The electric-field intensity at the $\zeta=0$ interface can be calculated from Eq. \eqref{eq7} as a function of both $F_d$ and $a$ (see below). Eq. \eqref{eq11} then results in a transcendental equation for $a$, which leads to the possible values of $a$ for a given value of $T = F_d$ that guarantee the fulfillment of the boundary conditions of the electric field at the interfaces of the slab. We want to emphasize that this procedure is slightly different but completely equivalent to those described in Refs. \cite{ChenPRB1987-1} and \cite{LeungPRB1989}.  In this regard, we have compared in Fig. \ref{fig2} the present results obtained by solving Eq. \eqref{eq11} with the calculations reported by Chen and Mills \cite{ChenPRB1987-1}. According to the definitions given in Ref. \cite {ChenPRB1987-1}, the transmission coefficient was obtained as a function of $\tilde{\lambda}=\frac{\alpha}{\varepsilon} \vert E_i \vert^2$ for $\varepsilon_0=\mu_0=1$ , $\mu=1$, and $\alpha>0$. The dashed line and the squares in Fig. \ref{fig2} correspond to a layer with $\varepsilon=4$, whereas the circles and the solid line correspond to a different layer with $\varepsilon=100$. In each case, as in Ref. \cite{ChenPRB1987-1}, we chose the thickness of the slab as $d = 0.4 \lambda_f$, where $\lambda_f = \frac{2 \pi}{\omega} \frac{c}{\sqrt{\varepsilon \mu}}$. Therefore $\zeta_d = 0.4 \frac{2 \pi}{\sqrt{\varepsilon \mu}}$ is independent of wave frequency. It is apparent from Fig. \ref{fig2}  that the present theoretical results are in good agreement with calculations by Chen and Mills \cite{ChenPRB1987-1}.

If we take $F_d = 1$, then the left-hand side of Eq. \eqref{eq11} vanishes for $F_0 = 1$. In other words, Eq. \eqref {eq11} is satisfied identically by simultaneously choosing $F_d = F_0 = 1$. In this situation, the incident electromagnetic field transmits completely through the slab. The complete transparency of the slab is due to the excitation of solitons within the nonlinear layer, that is, to the existence of electromagnetic states exhibiting unit transmission and maximum amplitude at the interfaces. From symmetry considerations, it can be seen that the function $F = F (\zeta)$ corresponding to a soliton state is an even function with respect to the center of the slab. The condition $F_d = F_0 = 1$ may be assumed as a necessary (but not sufficient) condition for the existence of solitons within the nonlinear layer.

The square $F$ of the electric-field amplitude may be obtained by integrating Eq. \eqref{eq7}, i.e, 
\be 
\label{eq12}
2(\zeta - \zeta_d) = \int_{F(\zeta_d)} ^{F(\zeta)} \frac{dF}{\sqrt{-\frac{a \mu}{2}}\sqrt{P(F)}},
\ee
where $P(F) = (F-F_1)(F-F_2)(F-F_3)$. The three roots of $P$ are given by
\be
\label{eq13}
F_1 = -\frac{1}{2} \left [ \sqrt{\left ( \frac{2p^2}{a \mu} + F_d \right )^2 + \frac{8\eta^2}{a\mu}F_d } + \frac{2p^2}{a\mu} + F_d \right ],
\ee
\be
\label{eq14}
F_2 = \frac{1}{2} \left [ \sqrt{\left ( \frac{2p^2}{a \mu} + F_d \right )^2 + \frac{8\eta^2}{a\mu}F_d } - \frac{2p^2}{a\mu} - F_d \right ],
\ee
and
\be
\label{eq15}
F_3 = F_d.
\ee
The solutions of Eq. \eqref{eq12} depend on the properties of the roots $F_1$, $F_2$, and $F_3$. As mentioned above, we have assumed that both $\varepsilon$ and $\mu$ are not frequency-dependent functions. Also, we have taken $\varepsilon>\mu>\varepsilon_0 = \mu_0=1$. For computational purposes we have taken $\varepsilon = 11.680$ and $\mu = 1.036$. If the hypothetical material layer we are considering exhibits a self-focusing nonlinearity ($a>0$) then the three roots of $P$ are real and $F_1<F_2<F_3$. In contrast, in the case of a self-defocusing nonlinearity ($a<0$) several situations of interest can occur. Let us define the quantities $\xi_1$, $\xi_2$, and $\xi_3$ such that
\be
\label{e16}
\xi_1 = -2 \left ( p^2 + 2 \eta^2 \right ) - 4 \eta \sqrt{p^2 + \eta^2},
\ee
\be
\label{e17}
\xi_2 = -2 \left ( p^2 + 2 \eta^2 \right ) + 4 \eta \sqrt{p^2 + \eta^2},
\ee
and
\be
\label{e18}
\xi_3 = \eta^2 - p^2,
\ee
respectively. One may note that $\xi_1 < \xi_2 < \xi_3 < 0$. The roots $F_1$ and $F_2$ are complex numbers whenever $\frac{\xi_1}{\mu F_d} < a < \frac{\xi_2}{\mu F_d}$. If $\frac{\xi_2}{\mu F_d} < a < 0$ or $a < \frac{\xi_1}{\mu F_d}$ then the three roots of $P$ are real. In the interval $\frac{\xi_3}{\mu F_d} < a < 0$ one has $F_1<F_3<F_2$, whereas if $\frac{\xi_2}{\mu F_d} < a < \frac{\xi_3}{\mu F_d}$ or $a < \frac{\xi_1}{\mu F_d}$ then $F_1<F_2<F_3$. In the above intervals Eq. \eqref{eq12} may be expressed in terms of the incomplete elliptic integral of the first kind \cite{Abramowitz}. Accordingly,
\be
\label{eq19}
F (\zeta) = F_2 + \left ( F_d - F_2 \right ) \mathrm{cn}^2 \left [ \kappa_f \left ( \zeta - \zeta_d \right ), m_f \right ] 
\ee
if $a>0$,
\be
\label{eq20}
F (\zeta) = \frac{F_1 + \sigma F_2 \mathrm{cn}^2 \left [ \kappa_{d1} \left ( \zeta - \zeta_d \right ), m_{d1} \right ] }{1 + \sigma \mathrm{cn}^2 \left [ \kappa_{d1} \left ( \zeta - \zeta_d \right ), m_{d1} \right ] }
\ee
within the interval $\frac{\xi_3}{\mu F_d} < a < 0$,
\be
\label{eq21}
F (\zeta) = \frac{F_d - F_2 \mathrm{sn}^2 \left [ \kappa_{d2} \left ( \zeta - \zeta_d \right ), m_{d2} \right ] }{1 - \mathrm{sn}^2 \left [ \kappa_{d2} \left ( \zeta - \zeta_d \right ), m_{d2} \right ] }
\ee
if $\frac{\xi_2}{\mu F_d} < a < \frac{\xi_3}{\mu F_d}$ or $a < \frac{\xi_1}{\mu F_d}$, and
\be
\label{eq22}
F (\zeta) = F_d + \Lambda \frac{1- \mathrm{cn} \left [ \kappa_{d3} \left ( \zeta - \zeta_d \right ), m_{d3} \right ] }{1 + \mathrm{cn} \left [ \kappa_{d3} \left ( \zeta - \zeta_d \right ), m_{d3} \right ] } 
\ee
if $\frac{\xi_1}{\mu F_d} < a < \frac{\xi_2}{\mu F_d}$. In the above expressions one has $\kappa_f = \sqrt{\frac{a \mu}{2} \left ( F_d - F_1 \right )}$, $m_f = \frac{F_d - F_2}{F_d - F_1}$, $\kappa_{d1} = \sqrt{\frac{\vert a \vert \mu}{2} \left ( F_2 - F_1 \right )}$, $m_{d1} = \frac{F_d - F_1}{F_2 - F_1}$, $\sigma = \frac{F_d - F_1}{F_2 - F_d}$, $\kappa_{d2} = \sqrt{\frac{\vert a \vert \mu}{2} \left ( F_d - F_1 \right )}$, $m_{d2} = \frac{F_2 - F_1}{F_d - F_1}$, $\Lambda = \sqrt{\frac{2}{\vert a \vert \mu} F_d \left ( \eta^2 - p^2 - a \mu F_d \right )}$, $\kappa_{d3} = \sqrt[4]{8 \vert a \vert \mu F_d \left ( \eta^2 - p^2 - a \mu F_d \right )}$, and $m_{d3} = \frac{1}{2} - \frac{1}{8} \frac{\delta}{\Lambda}$, where $\delta = \frac{4 p^2}{a \mu} + 6 F_d$. The behavior of the electromagnetic field around the points $a = \frac{\xi_i}{\mu F_d}$ ($i = 1$, 2, 3) can be studied separately. For example, from Eqs. \eqref{eq20} and \eqref{eq21} one obtains
\be
\label{eq23}
F(\zeta) \xrightarrow[a \rightarrow \frac{\xi_3}{\mu F_d}]{} F_d.
\ee
Similar analyzes can be performed in the cases $a \rightarrow \frac{\xi_1}{\mu F_d}$ and $a \rightarrow \frac{\xi_2}{\mu F_d}$ (results are not shown here).

\begin{figure}
\centering
\begin{tabular}{rr}
\epsfig{file=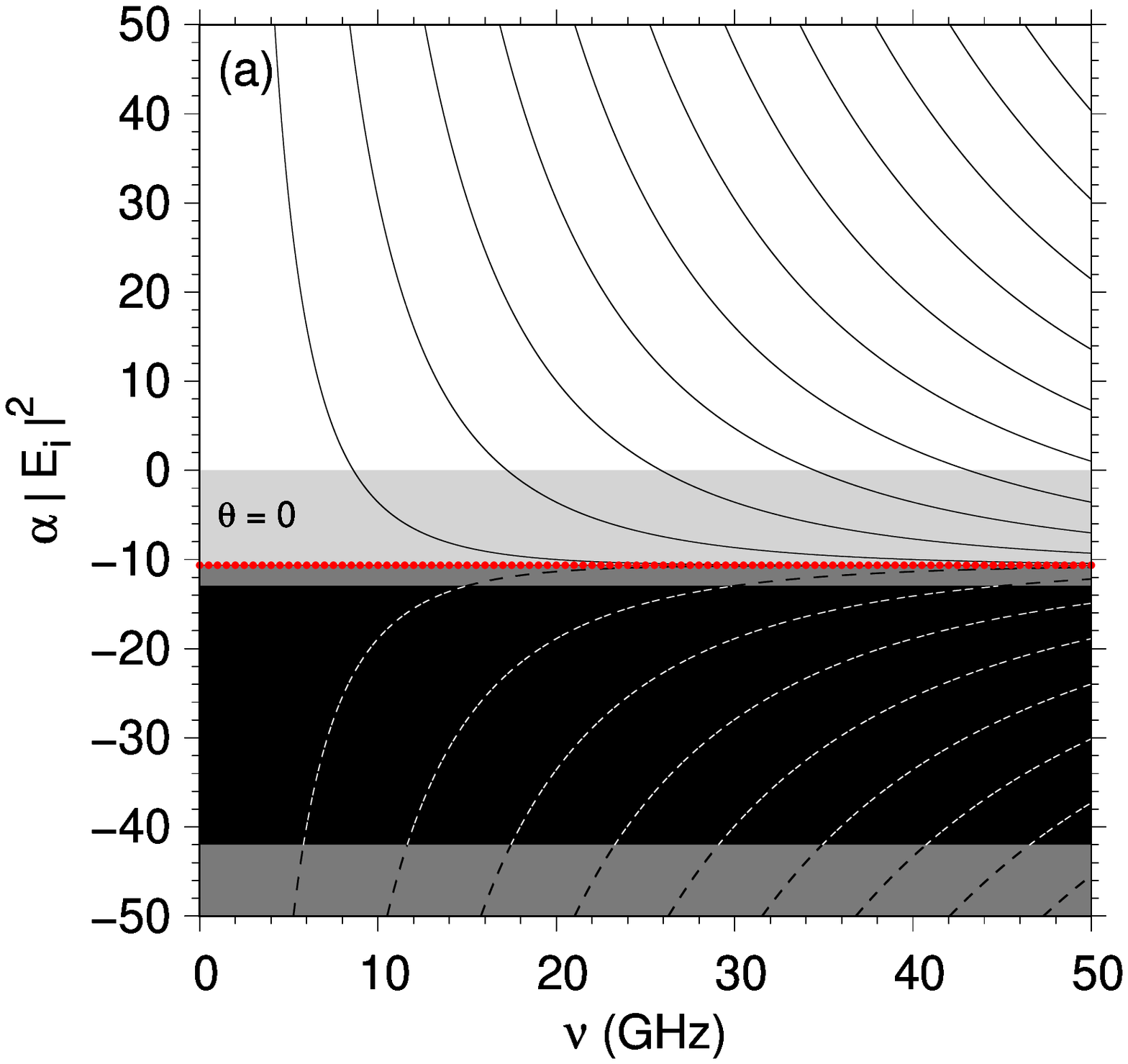, width=0.9\columnwidth}\crcr
\epsfig{file=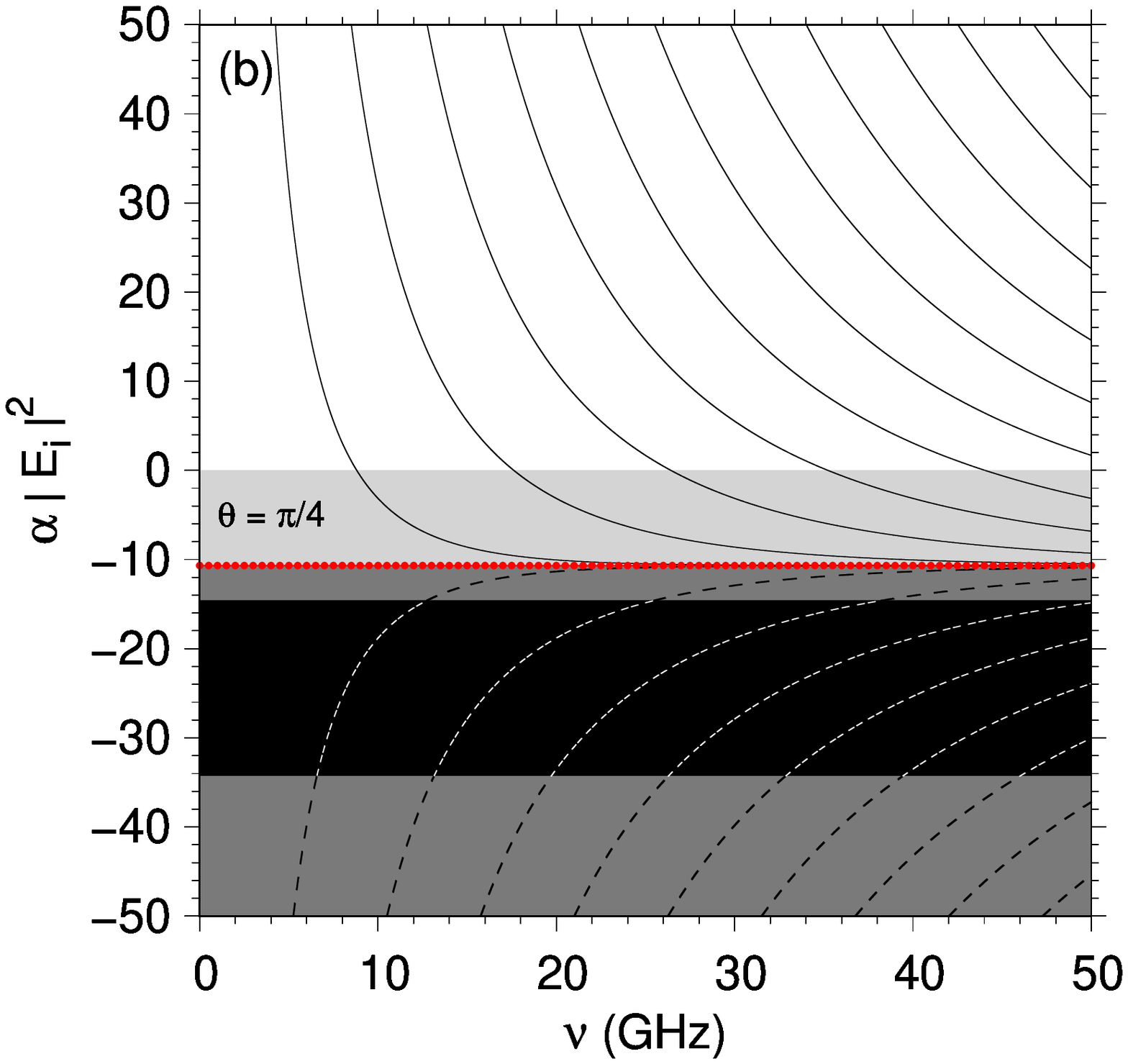, width=0.9\columnwidth}
\end{tabular}
\caption{(Color online) The solutions of Eqs. \eqref{eq24}-\eqref{eq25} in the $(\nu,a)$ space ($a = \alpha \vert E_i \vert^2$). Results were obtained for a nonlinear layer of width $d = 5$ mm. The white, light gray, upper gray, black, and lower gray regions correspond to $a>0$, $\frac{\xi_3}{\mu}<a<0$, $\frac{\xi_2}{\mu}<a<\frac{\xi_3}{\mu}$, $\frac{\xi_1}{\mu}<a<\frac{\xi_2}{\mu}$, and $a<\frac{\xi_1}{\mu}$, respectively. Panels (a) and (b) correspond to normal incidence and oblique incidence with $\theta= \frac{\pi}{4}$, respectively. Dotted lines correspond to $a = \frac{\xi_3}{\mu}$. Only the solid and dotted lines correspond to physically realizable soliton states (see the text).}
\label{fig3}
\end{figure}

\begin{figure}
\centering
\begin{tabular}{rr}
\epsfig{file=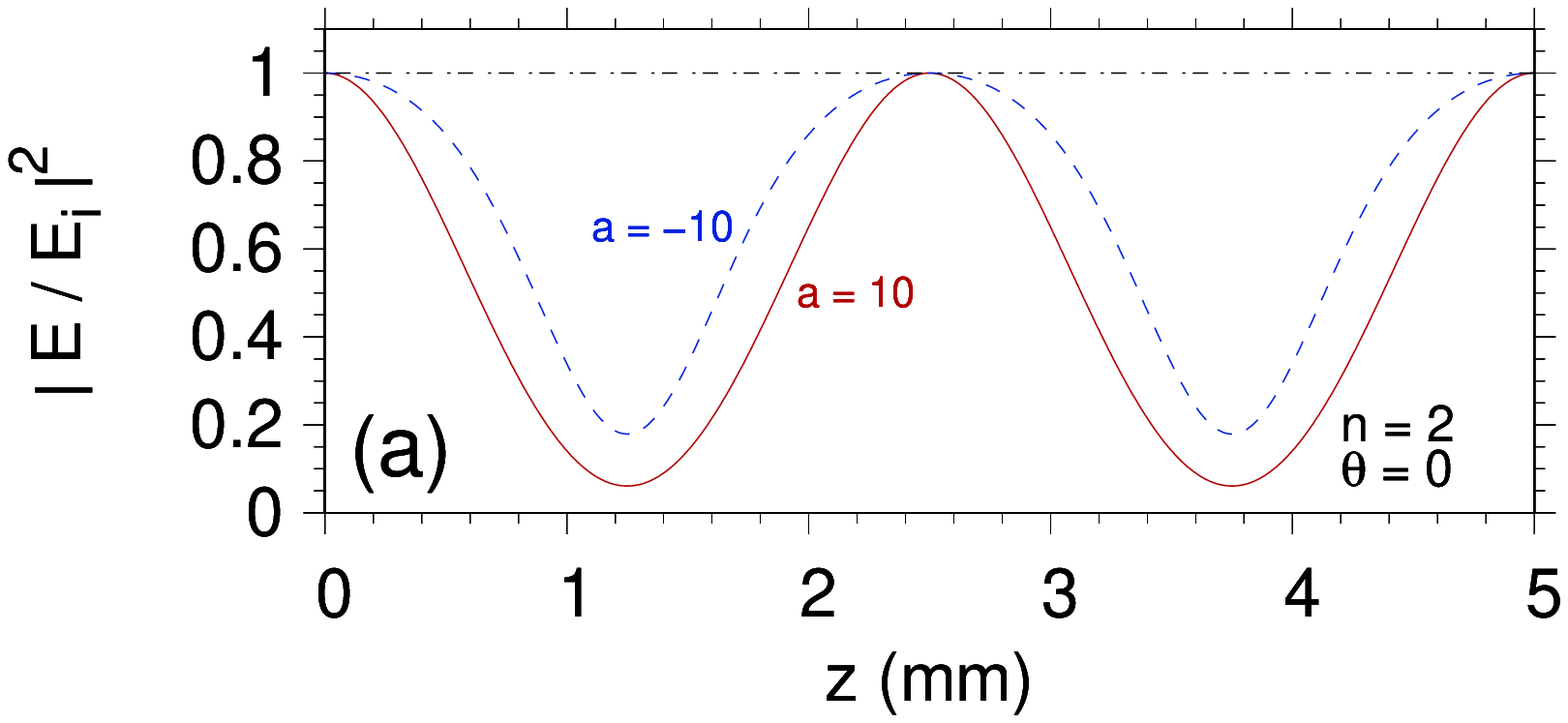, width=0.9\columnwidth}\crcr
\epsfig{file=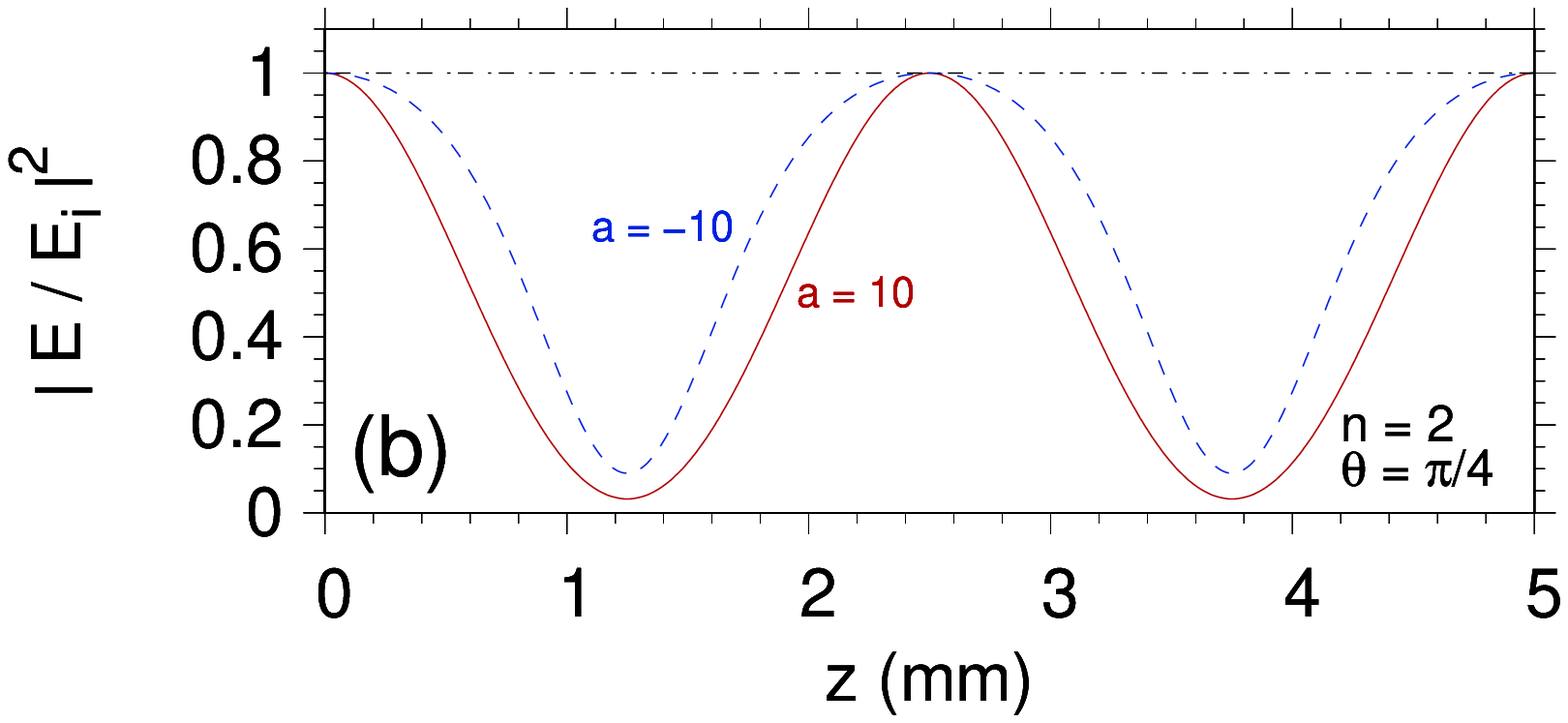, width=0.9\columnwidth}
\end{tabular}
\caption{(Color online) Square of the electric field amplitude as a function of the position within a nonlinear layer of thickness $d=5$ mm. Solid and dashed lines correspond to $a = 10$ and $a = -10$ within the intervals $a>0$ and $\frac{\xi_3}{\mu}<a<0$, respectively. Results were obtained for $n = 2$ (cf. the second solid line, from the left to the right, in both Figs. \ref{fig3}(a) and \ref{fig3}(b)), where the frequency values corresponding to $a = 10$ and $a = -10$ were obtained from Eqs. \eqref{eq24}-\eqref{eq25}. Panels (a) and (b) correspond to $\theta = 0$ and $\theta = \frac{\pi}{4}$, respectively. The dot-dashed horizontal lines correspond, in all panels, to the soliton solution obtained at the limit $ a \rightarrow \frac{\xi_3}{\mu}$. The same result is obtained regardless of the value of the wave frequency.}
\label{fig4}
\end{figure}

\begin{figure}
\centering
\begin{tabular}{rr}
\epsfig{file=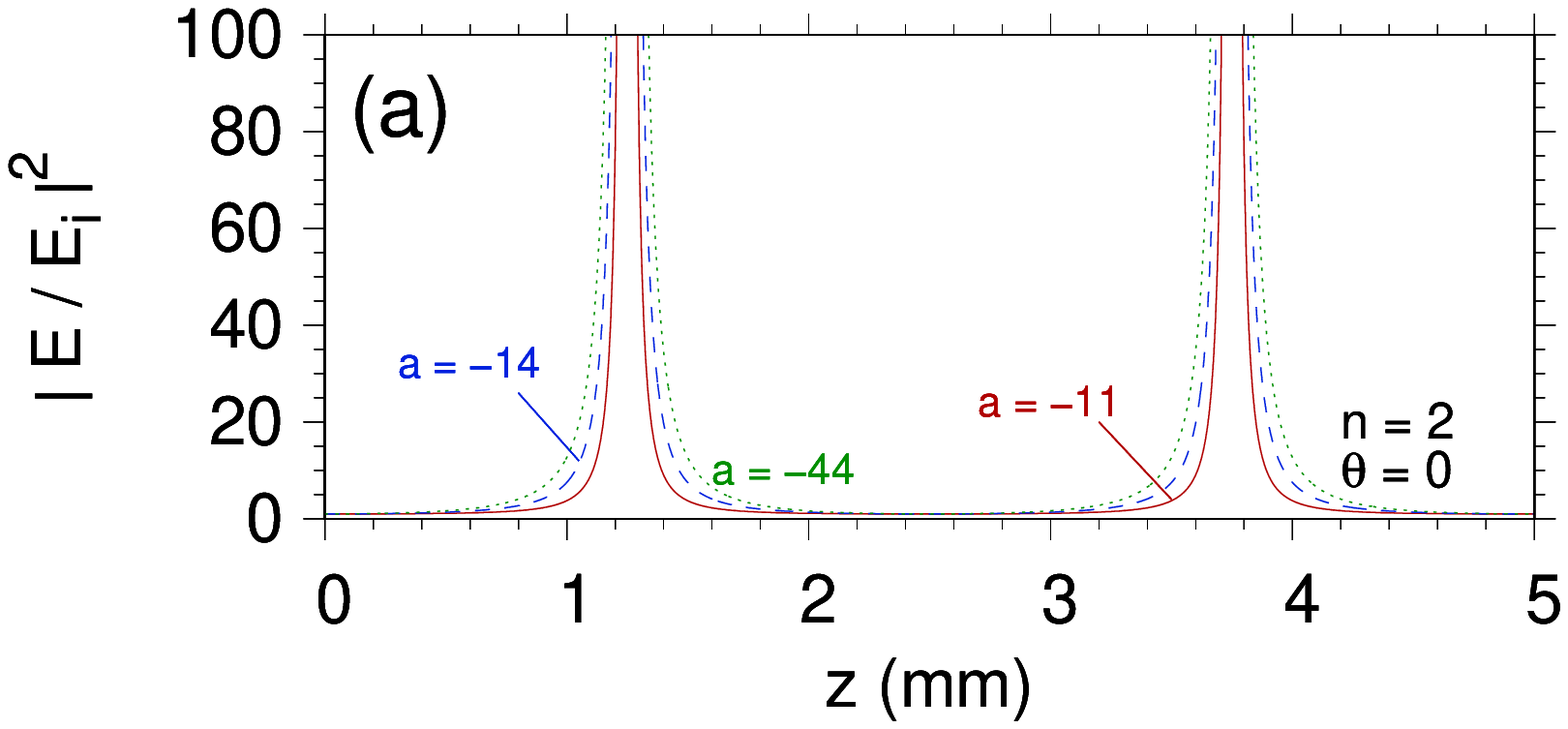, width=0.9\columnwidth}\crcr
\epsfig{file=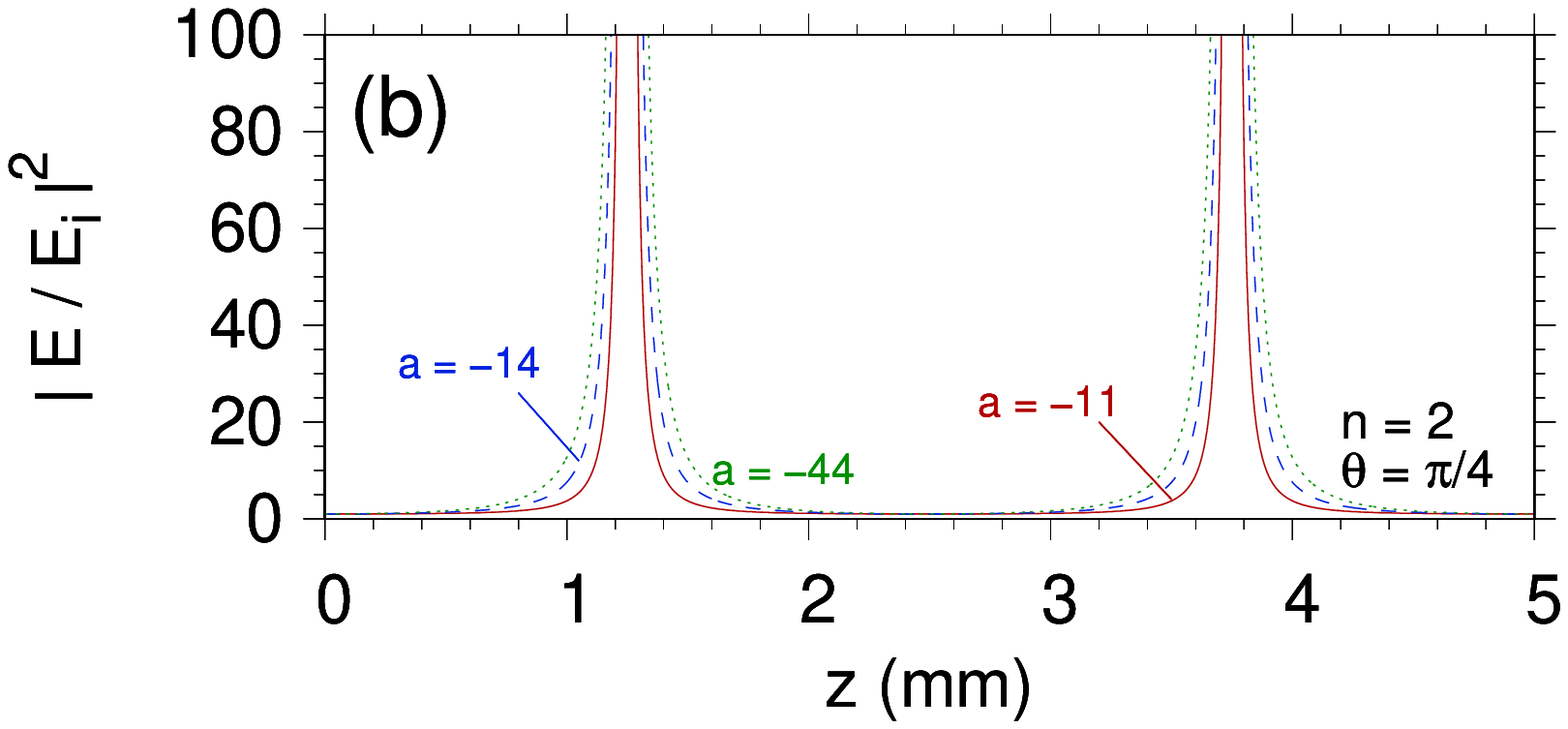, width=0.9\columnwidth}
\end{tabular}
\caption{(Color online) As in Fig. \ref{fig4}, but for solid, dashed, and dotted lines corresponding to $a = -11$, $a = -14$ and $a = -44$, respectively, within the regions $\frac{\xi_2}{\mu} <a<\frac{\xi_3}{\mu}$, $\frac{\xi_1}{\mu}<a<\frac{\xi_2}{\mu}$, and $a<\frac{\xi_3}{\mu}$, respectively. Results were obtained for $n = 2$ (cf. the second dashed line, from the left to the right, in both Figs. \ref{fig3}(a) and \ref{fig3}(b)). The corresponding frequency value for each value of $a$ was obtained from Eqs. \eqref{eq24}-\eqref{eq25}.}
\label{fig5}
\end{figure}

\begin{figure}
\centering
\epsfig{file=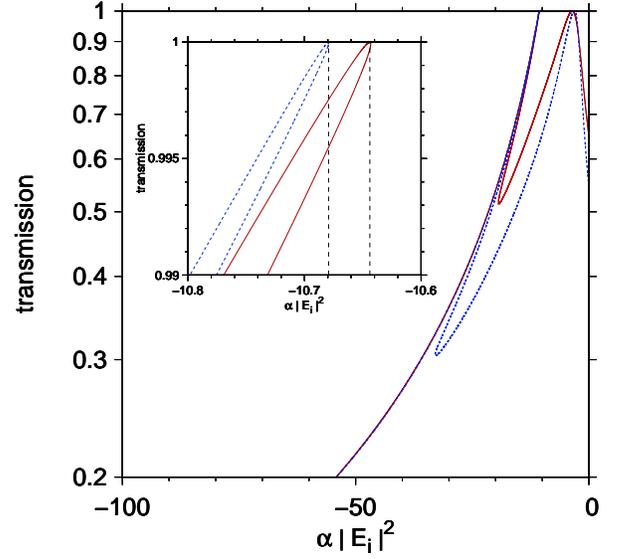, width=0.9\columnwidth}
\caption{(Color online) Transmission coefficient as a function of the normalized input intensity for $\nu = 10$ GHz. Results were obtained for a self-defocusing material with $d = 5$ mm. Solid and dashed lines correspond to $\theta = 0$ and $\theta = \frac{\pi}{4}$, respectively. Calculations were obtained by using the fourth order Runge-Kutta method. Vertical dashed lines in the inset indicate the position of the first soliton peak ($\alpha \vert E_i \vert^2 = \frac{\xi_3}{\mu}$) for each value of the incidence angle.}
\label{fig6}
\end{figure}

We can now study the existence of soliton states within the nonlinear layer by imposing the condition $F_d = F_0 = 1$ in Eqs. \eqref{eq19}-\eqref{eq22}. From such equations and by taking into account the properties of the Jacobian elliptic functions \cite{Abramowitz}, we find that
\be
\label{eq24}
\zeta_d = 2 n g(a),
\ee
where $n$ is a natural number and
\be
\label{eq25}
g(a) =
\begin{cases}
\frac{K(m_f)}{\kappa_f} & $if $ a>0 \cr
\frac{K(m_{d1})}{\kappa_{d1}} & $if $ \frac{\xi_3}{\mu} < a < 0 \cr
\frac{K(m_{d2})}{\kappa_{d2}} & $if $ \frac{\xi_2}{\mu} < a < \frac{\xi_3}{\mu} \cr
2 \frac{K(m_{d3})}{\kappa_{d3}} & $if $ \frac{\xi_1}{\mu} < a < \frac{\xi_2}{\mu} \cr
\frac{K(m_{d2})}{\kappa_{d2}} & $if $ a < \frac{\xi_1}{\mu} \cr
\end{cases}.
\ee
Eqs. \eqref{eq24} and \eqref{eq25} define a set of curves in the space $(\nu,a)$ at whose points the condition $F_d = F_0 = 1$ is fulfilled. In the above expression $K(m)$ is the complete elliptic integral of the first kind of parameter $m$ \cite{Abramowitz}. 

Fig. \ref{fig3} shows the numerical results obtained from Eqs. \eqref{eq24}-\eqref{eq25} for a nonlinear layer of thickness $d = 5$ mm. Calculations were performed for normal incidence and oblique incidence with $\theta= \frac{\pi}{4}$ (cf. Fig. \ref{fig3}(a) and Fig. \ref{fig3}(b), respectively). The results obtained for $a<0$ and $a>0$ were depicted on the same graph. Of course, the hypothetical nonlinear material considered here can only exhibit a self-defocusing or self-focusing nonlinearity. Therefore, present results must be understood in the sense of what would happen if our hypothetical material displayed one or another type of nonlinearity. In this sense, we highlight in Fig. \ref{fig3} the different regions of interest in the $(\nu, a)$ space with different gray scales. The white region corresponds to $a>0$. The light-gray region corresponds to the case $\frac{\xi_3}{\mu}<a<0$. In the upper gray region we have $\frac{\xi_2}{\mu}<a<\frac{\xi_3}{\mu}$, whereas $\frac{\xi_1}{\mu}<a<\frac{\xi_2}{\mu}$ in the black region and $a<\frac{\xi_1}{\mu}$ in the lower gray region.

As expected, when evaluating the function $F$ at each point of the solid curves shown in Fig. \ref{fig3} for $a>\frac{\xi_3}{\mu}$, the result satisfies the condition $F_0 = F_d = 1$. According to Eqs. \eqref{eq19}-\eqref{eq20}, the functions $F = F (\zeta)$ are, in these cases, continuous functions of $\zeta$ (or $z$) within the slab and are symmetric with respect to the center of the slab. Such electromagnetic states are soliton states propagating through the slab with $T=1$. Moreover, in the limit $a \rightarrow \frac{\xi_3}{\mu}$ [cf. Eq. \eqref{eq23} for $F_d=1$] an electromagnetic mode is obtained, which is spatially constant within the slab and does not depend on the wave frequency. As a consequence of Eq. \eqref{eq23}, such solution exhibits unit transmission if $F_d = 1$. In other words, the transmission coefficient tends toward one as $a \rightarrow \frac{\xi_3}{\mu}$, which results in a soliton state with the electric field uniformly distributed within the nonlinear layer regardless of the value of the wave frequency. Such soliton state was not discussed in previous works \cite{ChenPRB1987-1,LeungPRB1989}. The corresponding soliton lines in the $(\nu,a)$ space for $\theta=0$ and $\theta=\frac{\pi}{4}$ are shown in Figs. \ref {fig3}(a) and \ref{fig3}(b), respectively, as horizontal dotted lines on the boundaries between the respective light-gray and gray regions. We show in Fig. \ref{fig4}  the square of the electric field amplitude as a function of the position within the nonlinear layer of width $d=5$ mm. Calculations were performed for $n = 2$ and two different values of the incidence angle. Solid and dashed lines in Fig. \ref{fig4} correspond to $a = 10$ and $a = -10$, respectively, within the intervals $a>0$ and $\frac{\xi_3}{\mu}<a<0$, respectively, and are associated with the soliton states predicted by Eq. \eqref{eq25}. The dot-dashed horizontal line displayed in each panel of Figs. \ref{fig4} correspond to the soliton solution obtained at the limit $a \rightarrow \frac{\xi_3}{\mu}$. In each case, the electric-field amplitude is spatially constant within the slab, and the same result is obtained regardless of the value of the wave frequency, as explained above.

In contrast, the evaluation of $F$ at the points of the curves within the $(\nu, a)$ region such that $a<\frac{\xi_3}{\mu}$ (cf. dashed lines in Fig. \ref{fig3}) leads to solutions that are physically inadmissible. In spite of $F_d = F_0 = 1$ in these cases, the $F$ functions obtained from Eq. \eqref{eq21} or Eq. \eqref{eq22} are not continuous functions within the nonlinear layer. The singularities of $F$ can be characterized by performing a simple analysis. Let us suppose, for example, that $\frac{\xi_2}{\mu}<a<\frac{\xi_3}{\mu}$ or $a<\frac{\xi_1}{\mu}$. The poles of $F$ are obtained at the coordinate $\zeta$ satisfying the equation $\kappa_ {d2} (\zeta - \zeta_d) = -(2l-1) K(m_ {d2})$, where $l$ is a natural number. Furthermore, the condition $F_d = F_0 = 1$ leads to $\kappa_ {d2} \zeta_d = 2 n K(m_ {d2})$ (cf. Eq. \eqref {eq25}). By combining the last two equations, one has $\zeta \equiv \zeta_{nl} = \left [ 2 (n-l) + 1 \right ] \frac{K (m_ {d2})}{\kappa_ {d2}}$, or, equivalently,
\be
\label{eq26}
 z_{nl}=  \frac{c}{\omega} \zeta_{nl} = \frac{2(n-l)+1}{2n} \,d.
\ee
Since $0<z_{nl}<d$ one has $l=1$, 2, ..., $n$. Therefore, if $\frac{\xi_2} {\mu}<a<\frac{\xi_3} {\mu}$ or $a<\frac{\xi_1} {\mu}$, then the $n^{\mathrm{th}}$ electromagnetic mode that satisfies the condition $F_d = F_0 = 1$ will exhibit $n$ poles within the nonlinear layer. The $z_{nl}$ position of the $l^{\mathrm{th}}$ pole is given by Eq. \eqref{eq26}. In addition, it is possible to see that Eq. \eqref{eq26} is also valid within the region $\frac{\xi_1}{\mu}<a<\frac{\xi_2}{\mu}$, where $F$ is evaluated through Eq. \eqref{eq22}. We display in Fig. \ref{fig5} the existence of divergent solutions for the amplitude of the electric field as well as the position of their respective poles within the nonlinear layer of thickness $d=5$ mm. Calculations were performed for $n=2$, for two different values of the incidence angle, and for $a = -11$, $a = -14$ and $a = -44$ corresponding to the regions $\frac{\xi_2}{\mu} <a<\frac{\xi_3}{\mu}$, $\frac{\xi_1}{\mu}<a<\frac{\xi_2}{\mu}$, and $a<\frac{\xi_3}{\mu}$, respectively (cf. solid, dashed and dotted lines in Fig. \ref{fig5}, respectively).

A divergence of $F$ inside the nonlinear layer would imply that the nonlinear material could store infinite electromagnetic energy, a situation that is not observed in practice. This fact, together with the loss of continuity of $F$ under the condition $F_d = F_0 = 1$, indicates that the existence of soliton states is prohibited for each value of $a$ such that $a<\frac{\xi_3}{\mu}$. Consequently, the solutions of the equations \eqref{eq24}-\eqref{eq25} corresponding to such values of the normalized input intensity (see dashed lines in Fig. \ref{fig3}) are not physically realizable states. Furthermore, we would like to stress that our study on the properties of the electromagnetic field at the limits $a \rightarrow \frac{\xi_1}{\mu}$ and $a \rightarrow \frac{\xi_2}{\mu}$ did not lead to soliton states in these cases. 

The most important conclusion drawn from the present analysis is the existence of a limiting value of the normalized input intensity ($a = \frac{\xi_3}{\mu}$) below which soliton excitations in self-defocusing materials are not possible. In this regard, only the solid and dotted curves shown in Fig. \ref{fig3} represent the points in the $(\nu, a)$ space where soliton states can exist. We want to stress that the possibility of finding soliton excitations in a sufficiently wide frequency band essentially depends on the ordering of the roots $F_1$, $F_2$, and $F_3$ given by Eqs. \eqref{eq13}, \eqref{eq14}, and \eqref{eq15}, respectively. For nonlinear slabs with optical parameters dependent on the wave frequency, one could expect different possibilities in arranging the quantities $F_1$, $F_2$, and $F_3$, even in the case of self-focusing materials. In such a situation, the analysis of which regions of the frequency spectrum are allowed for soliton excitations could become a more intricate matter. Further works could be devoted to investigating this subject.

The existence of a limit for the excitation of soliton states in the self-defocusing material layer studied above can be verified by numerically solving Eq. \eqref{eq1}. In Fig. \ref {fig6} we show the transmission coefficient as a function of $a$ for $\nu = 10$ GHz. Calculations were obtained by using the fourth-order Runge-Kutta method. Solid and dashed lines correspond to $\theta = 0$ and $\theta = \frac{\pi}{4}$, respectively. In each case, the first peak from left to right corresponds to the soliton state obtained at $a = \frac{\xi_3}{\mu}$ (see the vertical dashed lines in the inset of Fig. \ref{fig6}). The second peak corresponds to a soliton state described by Eqs. \eqref{eq24}-\eqref{eq25}. From Fig. \ref{fig6} it follows that the transmission coefficient decreases monotonically as $a$ decreases, and no evidence of soliton excitation is observed in the region $a<\frac{\xi_3}{\mu}$.

In summary, we have theoretically investigated the excitation of solitons in optical layers exhibiting Kerr nonlinearities. We found the transmission coefficient as a function of the normalized input intensity $a = \alpha \vert E_i\vert^2$ corresponding to an incident monochromatic electromagnetic wave, which was assumed to be polarized according to the transverse electric configuration. Furthermore, the fundamental optical constants associated with the system were considered independent of the wave frequency. In the case of self-defocusing materials, the present study indicates that soliton excitations cannot occur {if the parameter $a$} is below a well-defined limiting value. We have shown that, at that limit, a soliton excitation can occur regardless of the value of the incident wave frequency. Present theoretical work extends and complements the previous results by Chen and Mills \cite{ChenPRB1987-1} and Leung \cite{LeungPRB1989}, and we hope that it will stimulate further experimental research on this topic.

\acknowledgments

The authors would like to thank the Scientific Colombian Agency CODI - University of Antioquia for partial financial support.


\end{document}